\documentstyle[aas2pp4]{article}

\lefthead{Iyomoto et al.}
\righthead{Activity Decline in NGC~1316 nucleus}

\begin{document}

\title{The Declined Activity in the Nucleus of NGC~1316} 
\author{Naoko Iyomoto\altaffilmark{1}, Kazuo Makishima\altaffilmark{1,2}, 
Makoto Tashiro\altaffilmark{1,2}, Susumu Inoue\altaffilmark{3},\\
Hidehiro Kaneda\altaffilmark{4}, Yukari Matsumoto\altaffilmark{1}, 
and Tsunefumi Mizuno\altaffilmark{1}}

\altaffiltext{1}{Department of Physics, School of Science, University of Tokyo,
7-3-1 Hongo, Bunkyo-ku, Tokyo 113-0033} 
\altaffiltext{2}{Research Center for the Early Universe, University of Tokyo, 
7-3-1 Hongo, Tokyo, Japan 113-0033}
\altaffiltext{3}{Department of Physics, School of Science, 
Tokyo Metropolitan University, 
1-1 Minami-Ohsawa, Hachioji, Tokyo, Japan 192-0364}
\altaffiltext{4}{The Institute of the Space and Astronautical Science, 
Yoshinodai, Sagamihara, Kanagawa, Japan 229-0022}

\begin{abstract}
NGC~1316 (Fornax~A) is a radio galaxy with prototypical double lobes,
where the magnetic field intensity is accurately measured 
via the inverse-Compton technique.
The radio-emitting electrons in the lobes are inferred to have 
a synchrotron life time of $\sim$ 0.1 Gyr.
Considering the lobe energetics,
we estimate the past nuclear X-ray luminosity of NGC~1316 to be at least
$\sim 4 \times 10^{34}$ W ($4 \times 10 ^{41}$ erg s$^{-1}$).
Thus, the nucleus was rather active at least 0.1 Gyr ago.
In contrast, we confirmed with {\it ASCA} and {\it ROSAT}
that the nucleus of NGC~1316 is very faint in X-rays at present, 
with the 2--10 keV luminosity of any AGN-like hard component 
being $< 2\times 10 ^{33}$ W ($2 \times 10 ^{40}$ erg s$^{-1}$)
even assuming a nuclear obscuration up to $\sim 10^{28}$ m$^{-2}$ 
(10$^{24}$ cm$^{-2}$).
This is at least an order of magnitude lower than the estimated past activity,
indicating that the nucleus is presently very inactive. 
From these two results, we conclude 
that the nucleus of NGC~1316 has become dormant during the last 0.1 Gyr.
This suggests the possible abundance of ``dormant'' quasars in nearby galaxies.
\end{abstract}


\section{Introduction}
Optical and X-ray deep surveys have revealed a clear evolution 
in active galactic nuclei (AGNs), that their co-moving space density 
was much higher in the past than at present (e.g. Hawkins et al. 1995).
Luminous AGNs are inferred to have turned into fainter AGNs 
or non-AGN objects (e.g. Padovani et al.1990), 
due probably to decrease on cosmological time scales
in the mass accretion rate onto the central black holes.

Recent radio (e.g. Miyoshi et al. 1995) and optical (e.g. Kormendy 1995) 
observations revealed a gaseous disk orbiting around a huge mass 
concentration, presumably a black hole, at the center of many nearby galaxies.
In addition, low-luminosity AGNs (LLAGNs), with 2--10 keV luminosities of 
$10^{32-34}$ W ($10^{39-41}$ erg s$^{-1}$), have been discovered with X-rays 
in many nearby galaxies (Makishima et al. 1994; Ptak et al. 1995; 
Ishisaki et al. 1996; Iyomoto et al. 1996, 1997; Terashima et al. 1998).
Therefore, relic of the past AGNs
may lurk in the center of many apparently normal galaxies.
Nevertheless, it has remained unclear whether 
these relics were long time ago actually luminous AGNs.

Here, we identify the
radio galaxy NGC~1316 (Fornax A) as a relic of a past luminous AGN. 
This is one of the brightest extra-galactic radio sources in the GHz band,
and exhibits double radio lobes (Ekers et al. 1983).
These lobes have allowed the first successful detection 
of inverse-Compton X-rays, by {\it ROSAT} (Feigelson et al. 1995) 
and {\it ASCA} (Kaneda et al. 1995; hereafter Paper I).
This in turm provides a reliable estimate of the magnetic field 
and synchrotron lifetime in the lobes.
By comparing the lifetime with a tight upper limit on the X-ray luminosity 
of its nucleus, we conclude that the activity of its nucleus has decreased 
markedly over the last $\sim 0.1$ Gyr.

We assume a distance of 16.9 Mpc to NGC~1316 (Tully 1988), 
corresponding to the Hubble constant of $\sim 75$ km s$^{-1}$ Mpc$^{-1}$.
All the errors refer to single-parameter 90\% confidence limits. 
We mainly use the MKSA units.

\section{Properties of the Nucleus of NGC~1316}
\subsection{Radio properties}
In contrast to the bright radio lobes, the nucleus (core) of NGC~1316
is quite radio faint; 0.026 Jy at 4.8 GHz.
Ratio of the core flux to the total (lobe + core) flux of NGC~1316 
becomes $3.6 \times 10^{-4}$ at 4.8 GHz, falling by two orders of magnitude
below the general correlation seen between these two quantities 
of radio galaxies (Morganti et al. 1993, 1995). 
This suggests that the NGC~1316 nucleus is currently rather inactive, 
and the lobes are relic of the past activity (e.g. Ekers et al. 1983). 
A supporting evidence is that the nucleus is offset by 
$\sim 6'$ ($\sim 30$ kpc) from the radio ``bridge'' connecting the two lobes.
The bridge may be the jet axis in the past, 
while the offset may represent the galaxy proper motion 
after the jet activity ceased (Ekers et al. 1983).

The radio emission, however, generally carries only a minor fraction of 
the overall radiative power emitted by an AGN,
and this fraction scatters largely among different types of AGNs.
For example, majority of Seyfert galaxies do not show 
luminous core radio emission, radio jets, or radio lobes.
Then, it might be that the NGC~1316 nucleus simply switched 
from a ``radio-loud'' mode to a ``radio-quiet'' mode,
rather than from active to inactive states. 
Alternatively the nucleus may in fact be quite active at present,
and the faintness of the radio nucleus may 
be a result of its obscuration from our line of sight
as suggested by the symmetric lobe morphology. 
Indeed, Fabbiano et al. (1994) 
reported a significant optical extinction in the NGC~1316 nucleus.

To obtain a firmer estimate of the current activity of the nucleus of NGC~1316,
we clearly need information in the higher-energy photons, particularly X-rays. 
Therefore, below we evaluate the X-ray activity of the NGC~1316 nucleus, 
utilizing the X-ray imaging data from {\it ROSAT} and 
the spectroscopic X-ray data from {\it ASCA}. 

\subsection{X-ray imagery with {\it ROSAT}}
Kim et al. (1998), using the {\it ROSAT} HRI,
reported that the X-ray nucleus of NGC~1316 is undetectable with
the 0.1--2.4 keV flux $<$ $1 \times 10^{-16}$ W m$^{-2}$.
In fact, the nucleus may be absorbed by hydrogen column density $N_{\rm H}$   
higher than $10^{26}$ m$^{-2}$ as in type II Seyfert galaxies 
(e.g. Iwasawa 1995; Ueno 1996; Turner et al. 1997),
although Kim et al. (1998) did not consider this.
Therefore in reference to the {\it ROSAT Users Handbook},
we converted their upper limit on the count rate 
into the intrinsic 2--10 keV luminosity of the nucleus, 
assuming a power-law spectrum of photon index 1.7
together with various values of $N_{\rm H}$.
These upper limits are plotted in Figure 1 as thin arrows, 
against the assumed absorption.
While the nuclear X-ray luminosity is well constrained in the low $N_{\rm H}$ range,
it is not constrained at $N_{\rm H}>10^{26}$ m$^{-2}$ 
due to the limited band pass of {\it ROSAT}.

\subsection{X-ray spectroscopy with {\it ASCA}}
The {\it ASCA} observation of NGC~1316 was made on 1994 January 11 for 35 ksec,
with the GIS (Gas Imaging Spectrometer; Ohashi et al. 1996, 
Makishima et al. 1996) in the normal PH mode and the SIS 
(Solid-State Imaging Spectrometer; Bruke et al. 1991) in the 2 CCD faint mode.
These data were already used in Paper I to detect the inverse-Compton X-rays. 
As shown in Figure 1b of Paper I, 
the X-ray emission from NGC~1316 itself dominates the GIS image,
and the X-ray centroid coincides with the optical nucleus 
within the astrometric accuracy of {\it ASCA} ($\sim 1'$).

We accumulated the GIS and SIS events over a circle of radii 4$'$ ($\sim$20 kpc) 
and 3$'$ ($\sim$15 kpc) centered on the X-ray centroid, respectively.
These regions cover the nucleus and a significant portion of the optical galaxy,
but excludes the radio lobes.
We added the data from the two GIS detectors (GIS2 and GIS3) 
into a single GIS spectrum, 
and the data from SIS0 chip 1 and SIS1 chip 3 into a single SIS spectrum.
The derived spectra are shown in Figure 2, 
after subtracting the background utilizing blank-sky observations.
They exhibit a soft component with low-energy emission lines,  
and a featureless hard continuum.
Both these components are characteristic of elliptical galaxies,
among which the former can be identified with
the thermal emission from hot interstellar medium 
(ISM; e.g. Forman et al. 1985; Canizares et al. 1987; Awaki et al. 1994).
The hard component is basically interpreted as
integrated hard X-rays from neutron-star binaries
(e.g. Makishima et al. 1989; Matsushita et al. 1994; Matsumoto et al. 1997),
but may be partially contributed by the active nucleus, 
because AGN emission and the integrated X-ray binary emission 
often exhibit similar spectral shapes in the {\it ASCA} energy band.

We fitted the GIS and SIS spectra jointly, 
with a spectral model consisting of a soft thermal plasma emission 
by Raymond \& Smith (1977) and a power-law. 
We applied a common absorption to the two components, with $N_{\rm H}$ 
fixed to the Galactic value of $1.5 \times 10^{24}$ m$^{-2}$ 
($1.5 \times 10^{20}$ cm$^{-2}$).
The fit is successful with $\chi^2 = 148$ for 140 degrees of freedom.
The Raymond-Smith component has a temperature of $0.77^{+0.03}_{-0.04}$ keV,
a metallicity of $0.11^{+0.06}_{-0.04}$ solar, 
and a luminosity of $(3 \pm 1) \times 10^{33}$ W in 0.5--4 keV.
These values are typical of the ISM emission from elliptical galaxies.
The power-law component has a photon index of $1.1 \pm 0.5$, 
and a 2--10 keV luminosity of 
$L_{\rm X}^{\rm hard} = (1.3^{+0.2}_{-0.3}) \times 10^{33}$ W.

Thus, the hard component would be consistent with 
an LLAGN emission in its spectral shape alone.
However, the value of $L_{\rm X}^{\rm hard}$
compares with the optical B-band luminosity of NGC~1316
($L_{\rm B}= 1.2 \times 10^{37}$ W at 16.9 Mpc; Tully 1988)
as $L_{\rm X}^{\rm hard}/L_{\rm B} = (1.0 \pm 0.2) \times 10^{-4}$.
This ratio aligns with the general X-ray to optical correlation 
seen among nearby normal galaxies without LLAGNs,
$L_{\rm X}^{\rm hard}/L_{\rm B} = (9.3 \pm 4.0) \times 10^{-5}$,
which was first derived with {\it Einstein} 
(e.g. Forman et al. 1985; Canizares et al. 1987)
and then updated using {\it Ginga} and {\it ASCA} (Makishima et al. 1989; 
Matsushita et al. 1994; Reynolds et al. 1997; Matsumoto et al. 1997).
Since the correlation is thought to represent the dominant contribution 
of neutron-star binaries to $L_{\rm X}^{\rm hard}$,
we infer that the hard component of NGC~1316 is also dominated 
by the integrated emission from neutron-star binaries, 
with little room for the LLAGN contribution.

Indeed, we can restore an acceptable spectral fit 
by replacing the power-law component 
with a Bremsstrahlung model of temperature fixed at 10 keV,
which simulates the binary-component spectrum   
(Makishima et al. 1989; Matsushita et al. 1994).
The value of $L_{\rm X}^{\rm hard}$ remained the same,
and the soft-component parameters did not change significantly.
Therefore, the hard component is consistent with 
being entirely of the binary origin.
Even allowing for $\pm 50$\% uncertainties in the 
$L_{\rm X}^{\rm hard}/L_{\rm B}$ correlation, 
the 2--10 keV luminosity of the NGC~1316 nucleus 
should not exceed $\sim 7.5 \times 10^{32}$ W. 
We show this limit in Figure 1 by an arrow with circle.

The above conclusion is insensitive to the nuclear absorption 
up to a column density of $N_{\rm H} \sim 10^{26}$ m$^{-2}$.
However the nucleus may be more heavily obscured, 
as seen in several LLAGNs (Makishima et al. 1994; Iyomoto et al. 1997).
Accordingly, we refitted the GIS and SIS spectra by a three-component model,
consisting of the Raymond-Smith component, the binary component,
and an absorbed power-law of photon index 1.7 simulating the absorbed LLAGN emission.
By fixing the column density of the power-law component at various values,
we calculated the maximum power-law normalization allowed by the data.
Thick arrows in Figure 1 represent the derived upper limits on the power-law luminosity, 
calculated in the 2--10 keV band after removing the absorption.
Thus, we can constrain the 2--10 keV luminosity of the nucleus as
$< 2 \times 10^{33}$ W over a wide range of nuclear obscuration up to 10$^{28}$ m$^{-2}$.

When the nuclear obscuration exceeds $N_{\rm H} \sim 10^{28}$ m$^{-2}$,
direct X-rays become almost invisible in the {\it ASCA} band.
However, we still expect to see scattered continuum, 
and a fluorescent Fe-K line (e.g. Makishima 1986).
Such Fe-K lines have been observed from many type II Seyferts
(e.g. Koyama et al. 1989; Ueno et al. 1994; Iwasawa et al. 1994; 
Fabian et al. 1994; Fukazawa et al. 1994; Iwasawa et al. 1997) 
and obscured LLAGNs (Iyomoto et al. 1997), 
at typical equivalent widths of 1--2 keV with respect to the scattered continuum.
Accordingly we fitted the 3--10 keV portion of the GIS/SIS spectra
with a single power-law plus a narrow Gaussian line
of which the center energy is fixed at 6.4 keV (neutral Fe-K line). 
Then the Fe-K line equivalent width has been constrained to be $< 280$ eV.
Therefore, the scattered continuum accounts for 
at most $\sim 1/7$ of the observed hard X-ray component.
By correcting this for the scattering efficiency,
we can constrain the intrinsic nuclear luminosity.
Although the scattering efficiency depends 
on geometry and ionization state of the scatterer,
we may adopt a rough estimate of $\sim$1\% at 6 keV (e.g. Koyama et al. 1989),
to crudely constrain the intrinsic luminosity as $< 1.4 \times 10^{34}$ W.
This limit is also given in Figure 1 by an arrow with square.

\subsection{Other properties}
We may quote further evidences supporting 
the current dormancy of the NGC~1316 nucleus. 
The [OIII] luminosity of NGC~1316 is lower than 
$3 \times 10^{31}$ W (at 16.9 Mpc; Siebert et al. 1996),
which corresponds to a 2--10 keV luminosity of $< 5 \times 10^{33}$ W
in view of the tight X-ray to [OIII]$\lambda$5007 correlation 
in type I Seyferts (Grossan 1992; Mulchaey et al. 1994).
The amounts of molecular and neutral hydrogen in NGC~1316
are estimated to be 8.95 $M_{\odot}$ using SCO line,
and $<8.87~M_{\odot}$ using 21 cm line, respectively.
These are no higher than those in normal E/S0 galaxies (Roberts et al. 1991),
indicating that the nucleus lacks sufficient fueling material.
Infrared flux densities of NGC~1316, 0.25 Jy at 25 $\mu$m and 3.0 Jy at 60 $\mu$m, 
are similar to or even lower than those of normal E/S0 galaxies (Green et al. 1992).
This rules out the possibility 
that the X-rays from a luminous AGN are 
converted into infrared by the obscuring material.
All these properties support the conclusion from the radio and X-ray observations,
that the nucleus is in the state of very low activity.

\section{Properties of the Lobes of NGC~1316}
\subsection{Lobe lifetimes}
While the nucleus of NGC~1316 is currently nearly dormant as discussed in \S 2, 
it must have been much more active previously
because it had to power the lobes. Then, when did the activity decline?

The inverse-Compton X-rays from NGC~1316 are produced 
when the radio-emitting relativistic electrons in the lobes 
scatter off the cosmic microwave photons.
Comparing the radio and X-ray brightness, 
the magnetic field strength in the lobes of NGC~1316 has been 
determined unambiguously to be 0.3 nT (Feigelson et al. 1995; Paper I),
without relying upon the usual assumption of energy equipartition.
In addition, the lobe radio spectrum is known to extend at least to 5 GHz 
without a noticeable spectral turn over (Brinkmann et al. 1994 [408 MHz]; 
Jones et al. 1992 [843 MHz]; Ekers et al. 1983 [1415 MHz]; 
Wall et al. 1973 [2700 MHz]; Morganti et al. 1993 [4.8 GHz]). 
Then, it is a textbook case to calculate 
that the electrons emitting 5 GHz radio photons lose half the energy 
on a characteristic time scale of $\tau_1 \sim 0.08$ Gyr.
This $\tau_1$ gives a reliable measure of the 
synchrotron life time of the lobes of NGC~1316.
That is, the nucleus must have been active until
at least $\tau_1$ years ago, or even to a more recent epoch.

The $6'$ (30 kpc) offset between the nucleus and the bridge provides another clue.
Since NGC~1316 is a member of the Fornax cluster 
of which the velocity dispersion is
$\sim 300$ km s$^{-1}$ (Ferguson \& Sandage 1990), 
the time required for the nucleus to travel this distance may be 
$\tau_2 \sim 0.1$ Gyr (30 kpc divided by  $\sim 300$ km s$^{-1}$).
This must be a lower limit on the time elapsed since the jet ceased, 
as the bridge must have a non-zero lifetime. 

The first of the above two arguments suggests that the activity decline time scale 
$\tau$ is shorter than $\tau_1 \sim$ 0.08 Gyr,
while the second one requires $\tau$ to be longer than $\tau_2 \sim$ 0.1 Gyr.
To satisfy both these conditions, $\tau \sim 0.1$ Gyr is required.

\subsection{Kinetic power of the lobe sustainment}
The remaining task is to estimate how powerful the nucleus used to be.
For this purpose, we note that the electron energy density in the lobes 
have been constrained in Paper I to be $(3.0 \pm 1.3) \eta^{-1}
(D/20 {\rm Mpc})^{-1} \times 10^{-14}$ J m$^{-3}$, where $D$ is the distance 
and $\eta <1$ is volume filling factor of the relativistic electrons.
By multiplying this with the lobe volume of $\sim 1.5 \times 10^{64}$ m$^{3}$,
we estimate the electron energy contained in the lobes to be 
$U_{\rm e} \sim 1.0 \times 10^{51}$ J (assuming $\eta \sim$ 1).
Since this energy is radiated away on the synchrotron lifetime of 
$\tau_1=0.08$ Gyr, a kinetic power of at least 
$U_{\rm e}/ \tau_1 \sim 4 \times 10^{35}$ W is needed to sustain the lobes 
in a steady-state condition.

There is a good correlation between kinetic luminosities
and narrow line luminosities of AGNs over five orders of magnitude
(Rawlings \& Saunders 1991; Ceolotti \& Fabian 1993).
The luminosity of [OIII], the strongest narrow emission line, 
in turn correlates well with the X-ray continuum luminosity 
over five orders of magnitude (Grossan 1992; Mulchaey et al. 1994).
From these two correlations, we infer that the nucleus of NGC~1316 
used to have a 2--10 keV luminosity of $4 \times $10$^{34}$ W or larger.
However we admit that the above two correlations are 
both subject to significant intrinsic scatters,
so that the derived X-ray luminosity is a very crude estimate. 
We may rather argue that typical radio galaxies have 0.5--4 keV 
luminosities (to which the 2--10 keV luminosities should be comparable)
in the range $10^{34-37}$ W (Fabbiano et al. 1984),
and none of the LLAGNs mentioned in \S1 exhibits prominent radio lobes.

\section{Discussion and Conclusion}
Using the {\it ASCA} data, we have shown that the nucleus of NGC~1316 is 
at present very inactive, with the 2--10 keV luminosity less than 
$2 \times 10^{33}$ W ($2 \times 10^{40}$ erg s$^{-1}$) 
over a wide range of obscuration up to 10$^{28}$ m$^{-2}$.
This agrees with the extreme faintness of the radio core.
In contrast, the nucleus must have been previously as luminous as 
$> 4 \times 10^{34}$ W in the 2--10 keV X-rays.
These indicate that the mass accretion rate onto the central black hole of 
NGC~1316 decreased by more than one order of magnitude.
Our X-ray constraints become rather loose
for nuclear obscuration higher than $N_{\rm H} \sim 10^{28}$ m$^{-2}$,
but the available data in other wavelengths altogether suggest
that such a heavy obscuration is rather unlikely.

Furthermore, using the estimate of the lobe life times 
and the observed nucleus vs. bridge offset, we have shown 
that the activity decline took place in the last $\tau \sim 0.1$ Gyr.
These results reveal a large decrease in the AGN activity of NGC~1316, 
and provide one of the first convincing examples 
that the activity of an AGN can actually turn-off.

Objets like NGC~1316 must be statistically rather rare, because its turn-off 
took place over a time scale which is much shorter than the Hubble time.
Its lobes are currently disappearing, and in another 0.1 Gyr or so, 
NGC~1316 will be a quite normal elliptical galaxy without any trace of 
the past AGN activity.
This suggests that a certain fraction of currently normal galaxies used to be 
luminous AGNs in the past, although quantitative estimates are difficult 
with only one example.
In conclusion, the fascinating case of NGC~1316 supports the scenario 
that the past quasars turned into normal galaxies.

\vspace{5mm}
We thank the members of the {\it ASCA} team.
This work was supported in part by the Grant-in-Aid for Scientific Research
(07CE2002) of the Ministry of Education, Science and Culture in Japan.

\figcaption{Upper limits on current nuclear luminosity of NGC~1316 in the energy range of 2--10 keV as a function of assumed absorption. They are shown after removing the absorption. Thin and thick arrows represent those estimated with {\it ROSAT} and {\it ASCA}, respectively. An arrow with circle shows that estimated assuming no intrinsic absorption. An arrow with square is that estimated from the upper limit on equivalent width of the fluorescent Fe-K line, assuming complete obscuration of direct nuclear X-ray emission. \label{fig1}}

\figcaption{(a) The GIS spectrum of NGC~1316 integrated within a circle of 4$'$ radius around X-ray brightness peak. The upper panel shows the data with the best fit model of the two component fit after subtracting the background but without removing the instrumental response. The lower panel shows the residuals. (b) Same for (a), but derived with the SIS within a circle of 3$'$ radius. \label{fig2}}

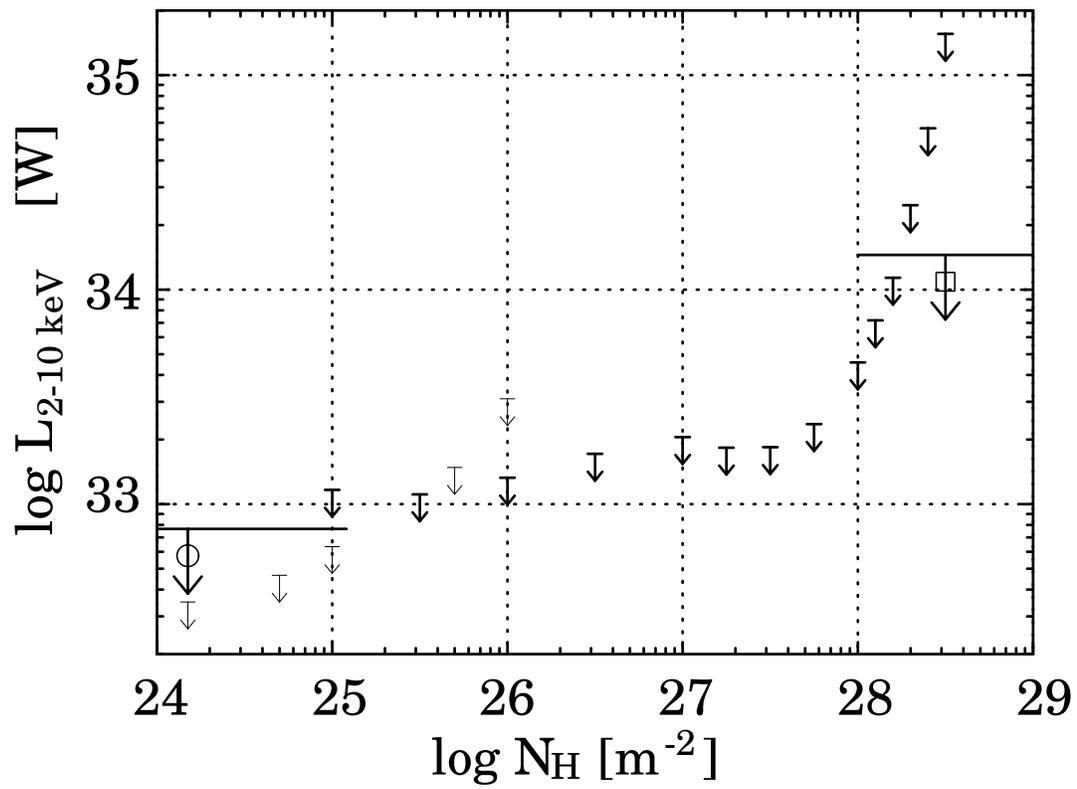
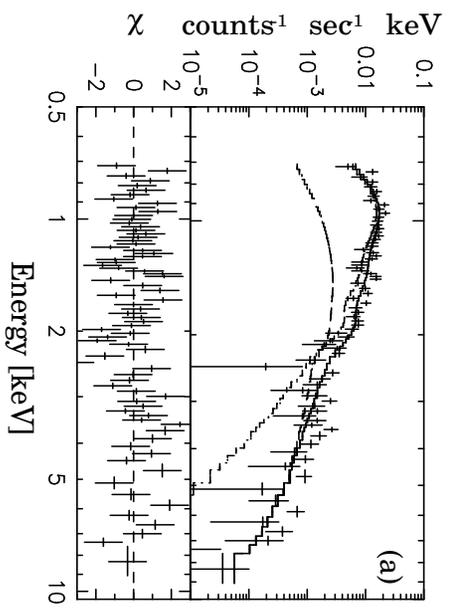
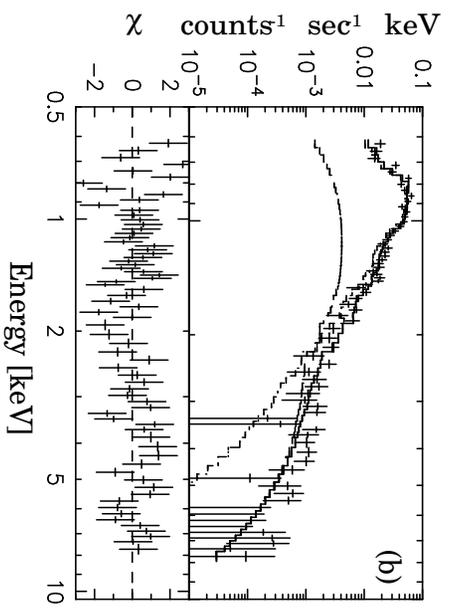
\end{document}